# Virtual Lung Screening Trial (VLST): An *In Silico* Study Inspired by the National Lung Screening Trial for Lung Cancer Detection

Fakrul Islam Tushar[1,2], Liesbeth Vancoillie[1], Cindy McCabe[1,3], Amareswararao Kavuri[1], Lavsen Dahal[1,2], Brian Harrawood[1], Milo Fryling[1], Mojtaba Zarei[1,2], Saman Sotoudeh-Paima[1,2], Fong Chi Ho[1,2], Dhrubajyoti Ghosh[4], Michael R. Harowicz[1], Tina D. Tailor[1], Sheng Luo[4], W. Paul Segars[1,2,3], Ehsan Abadi[1,2,3], Kyle J. Lafata[1,2,3,5], Joseph Y. Lo[1,2,3*], Ehsan Samei[1,2,3*]

[1] Center for Virtual Imaging Trials, Carl E. Ravin Advanced Imaging Laboratories, Dept. of Radiology, Duke University School of Medicine

[2] Dept. of Electrical & Computer Engineering, Pratt School of Engineering, Duke University

[3] Medical Physics Graduate Program, Duke University

[4] Dept. of Biostatistics and Bioinformatics, Duke University School of Medicine

[5] Dept. of Radiation Oncology, Duke University School of Medicine.

[*] These authors contributed equally as co-senior authors.

## Abstract

Clinical imaging trials play a crucial role in advancing medical innovation but are often costly, inefficient, and ethically constrained. Virtual Imaging Trials (VITs) present a solution by simulating clinical trial components in a controlled, risk-free environment. The Virtual Lung Screening Trial (VLST), an *in silico* study inspired by the National Lung Screening Trial (NLST), illustrates the potential of VITs to expedite clinical trials, minimize risks to participants, and promote optimal use of imaging technologies in healthcare. This study aimed to show that a virtual imaging trial platform could investigate some key elements of a major clinical trial, specifically the NLST, which compared Computed tomography (CT) and chest radiography (CXR) for lung cancer screening. With simulated cancerous lung nodules, a virtual patient cohort of 294 subjects was created using XCAT human models. Each virtual patient underwent both CT and CXR imaging, with deep learning models, the AI CT-Reader and AI CXR-Reader, acting as virtual readers to perform recall patients with suspicion of lung cancer. The primary outcome was the difference in diagnostic performance between CT and CXR, measured by the Area Under the Curve (AUC). The AI CT-Reader showed superior diagnostic accuracy, achieving an AUC of 0.92 (95% CI: 0.90-0.95) compared to the AI CXR-Reader's AUC of 0.72 (95% CI: 0.67-0.77). Furthermore, at the same 94% CT sensitivity reported by the NLST, the VLST specificity of 73% was similar to the NLST specificity of 73.4%. This CT performance highlights the potential of VITs to replicate certain aspects of clinical trials effectively, paving the way toward a safe and efficient method for advancing imaging-based diagnostics.

## 1. Introduction

Lung cancer ranks as the leading cause of cancer-related deaths, accounting for approximately 1.8 million fatalities in 2020.[1,2] Projections from the American Cancer Society indicate that an estimated 238K individuals in the United States are anticipated to be diagnosed with lung cancer in 2023.[3] In the realm of timely detection and diagnosis of lung cancer, imaging modalities like chest radiography (CXR) and Computed Tomography (CT) scans play a crucial role for not only the diagnosis of lung cancer but also those of a wide range of abnormalities.[4]

Associating the early-state cancer detection to the larger likelihood of cure, multiple lung cancer screening trials worldwide have contributed valuable insights into the efficacy of lung cancer screening, such as National Lung Screening Trial (NLST),[4] Nederlands-Leuvens Longkanker Screenings Onderzoek (NELSON),[5] Multicentric Italian Lung Detection (MILD),[6] British Thoracic Society Lung Cancer Screening Group (BTSLSG),[7] and International Early Lung Cancer Action Program (IE-LCAP) trial.[8]

In spite of their benefits, clinical trials are inefficient and costly. Over its duration of 8 years, for example, NLST enrolled over 50,000 subjects at a cost of $256 million.[4,9] At the conclusion of such trials, the diagnosis or treatment being evaluated may already be obsolete. For medical imaging, that risk is particularly high because technology evolves so rapidly. Additionally, clinical trials may involve putting patients at risk, including that of radiation exposure or overdiagnosis leading to unnecessary procedures. Therefore, it is imperative to develop alternative trial procedures that can address these gaps to ensure financial feasibility, timely change of clinical practice, and patient safety.[10]

Virtual Imaging Trials (VITs) leverage advances in computational techniques to simulate the workflow of clinical imaging trials, i.e., patients undergoing scans that are interpreted by readers. Compared to clinical trials, VITs may serve as an alternative that is faster, safer, and more cost-effective. In their comprehensive review paper, Abadi *et al.* delve extensively into VITs using numerous imaging modalities to study myriad applications, such as for coronavirus disease (COVID-19), emphysema, and organ dosimetry.[11] There are still very few VIT studies that emulate an end-to-end trial, however, due to the immense complexity of simulating all trial aspects. The most mature VIT studies to date have focused on breast cancer screening, such as the trailblazing study by FDA scientists to emulate a comparative clinical trial of mammography versus breast tomosynthesis.[12]. To advance VIT applications, simulation software modules need to be integrated to facilitate research that is both expedient and rigorous, and the trials need to expand to other clinical settings with high clinical impact.

To address these gaps, this Virtual Lung Screening Trial (VLST) study aimed to take the first steps toward an "*in silico*" emulation to compare efficacy of CXR and CT in lung cancer screening. The NLST was extremely important in medical imaging as it changed clinical practice and led to the adoption of current lung cancer screening programs. Inspired by that trial, our VLST study presents three primary advances. First, a virtual patient cohort was designed to encompass the many anatomical structures of the chest and multiple types of lung nodules. Second, this study implemented virtual scanners for CXR and CT, modalities that have attracted considerable recent attention in machine learning research due to their high clinical volume and impact.[13,14] Finally, image interpretation was performed by virtual readers implemented using reproducible deep learning models. By integrating these software modules, this study presented the first end-to-end virtual imaging trial in the new domain of chest imaging, designed to evaluate an ***in silico*** platform for a task with high clinical relevance.

## 2. Methods

### 2.1. Study Design

**Fig. 1** shows an overview of the steps performed to accomplish the VLST consisted of simulating a virtual population, modeling virtual scanners, developing virtual readers, and finally analyzing the detection and diagnosis performance. In the section below, we will briefly explain each of these steps.

### 2.2. Virtual Population

In any clinical trial, the selection of the targeted population and sample groups is of utmost importance, as it directly influences the validity of trial results and outcomes. The creation of a virtual human subject involved two key steps: constructing the normal anatomy followed by insertion of the specific abnormalities. For the VLST, the subjects employed were computational human models generated from full body CT scans from a single health system (Duke University Health System) which encompasses multiple hospitals. The construction of these human models involved a series of steps: beginning with the segmentation of specific organs, ensuring the quality control of these segmentations, followed by the creation of airways and vessels, and culminating in the voxelization process. An outline of this procedure is shown in **Fig. 1**, with further information available in our previous publication.[15] In this research, we employed both pre-existing and newly developed extended cardiac-torso (XCAT) models representing both sexes with varying age, BMI, and race.[15-17] Due to computational constraints to generate the complex 3D dataset, this initial study was limited to only 294 virtual subjects. Due to the small sample size, which is discussed in the Limitations section of the discussion, future work will leverage enhanced resources to expand the cohort. Demographically, the mean age was 59 years, with a distribution between 145 men, 116 women, 33 unknown from older phantom participants.

Simulated nodules were generated in two steps. First, a single-density lesion was formed, adhering to the desired morphology. Subsequently, a convolution process created a multi-density structure using a previously published approach. [18,19] Several instances of these simulated nodules, featuring diverse characteristics such as size and morphology, are depicted in **Fig. A.1 and Fig. A.2**. Consistent with methods used in previous studies, simulated lesions were designed to represent cancerous nodules.[18,19] These simulated nodules were then randomly inserted into the lungs of 174 patients, guided by indications of where nodules have been observed in prior clinical trial studies.[4] A total of 512 solid lesions were created including 202 (39.4%) with homogeneous texture and 310 (60.5%) heterogeneous texture. The remaining 120 patients were designed without any nodules. The ratio of individuals with and without nodules reflects the combination of pre-existing and newly created virtual subjects while ensuring a greatly enriched prevalence of >50% cancer to facilitate analysis of reader performance. Specifics regarding population demographics, along with inclusion and exclusion criteria, are clarified in **Table 1** and **Fig. 2**.

### 2.3. Virtual Scanners and Imaging Protocols

The cohort of computational human models with and without nodules was virtually imaged using a validated imaging simulation tool DukeSim.[20-22] DukeSim generates projection images from voxelized computational models using ray tracing (for primary signal) and Monte Carlo simulation (for scatter signal and radiation dose). Both modalities were designed to align with the acquisition technologies used during the NLST era, ensuring consistency in approach. To create CXR images, raw projection images were post-processed to replicate the contrast, noise, and resolution of the NLST-era images. For CT, projections were reconstructed using the vendor-neutral reconstruction tool MCR Toolkit[23] with weighted filter back projection configured to mimic the physical and geometrical characteristics of two generic CT scanners, named Duke Legacy W12 and Duke Legacy W20, which are representative of NLST CT systems.[24] The geometry and acquisition configurations of the scanners are listed in **Appendix Table B.1**. Samples of human model and simulated images with lesions are presented in **Fig. 3**.

### 2.4. Virtual Reader

The virtual reader component of our study deployed deep learning models to emulate the image interpretation by radiologists. Two RetinaNet models were developed: a 2D model for CXR images and a 3D model for CT volumes.[25] This readily available machine learning architecture was trained with publicly accessible clinical datasets (LUNA16,[26] NODE2[27]) to ensure ease of replication and minimize virtual reader variability across a variety of diagnostic scenarios for both modalities. The architecture, training methodologies, and patch extraction techniques remained constant across both models, with only the training datasets and data dimensions varying. This approach is akin to having the same radiologist interpret different modalities with consistent training.

The workflow started with image data augmentation and utilized RetinaNet's feature pyramid network to extract multi-scale features.[25] Anchor boxes generated were matched with reference standard annotations provided with each public dataset, refined through regression, and classified to detect actionable nodules.[28] Then for the VLST study, the simulated CXR and CT image data were used for independent, external validation. For each virtual patient, each model detected the candidate locations and corresponding probabilities of being actionable nodules. Additionally, for each patient without nodules, one random location within the lungs was also analyzed by the model to represent negative backgrounds. Detailed information regarding the development, validation, and clinical datasets used for the virtual reader models is documented in **Appendix C**.

### 2.5. Trial End Point

Since each patient exam may receive multiple detections, the lesion with the highest probability value among all lesions in the patient was chosen to represent the aggregated, overall diagnosis for each patient. This patient-level performance replicated the decision-making process of radiologists who search for initial lesion candidates, then use the most suspicious as the index lesion to inform follow-up recommendations. The VLST's primary endpoint was the difference between CT and CXR for the recall for lung cancers at the initial baseline screening exam. Additionally, subgroup analyses were conducted to evaluate the performance difference for the two distinct lesion texture types. In contrast, clinical screening trials like NLST had a much more complex design involving three annual screening rounds and the final endpoint was based on mortality outcomes years later. Reproducing those complexities is beyond current VIT approaches, so the VLST endpoint was focused only on the reader performance for the cancer recall task, reflecting an earlier stage of clinical decision-making.

### 2.6. Statistical Analysis

Performance of VLST models was assessed using the Receiver Operating Characteristic (ROC) Area Under the Curve (AUC) at the patient level, with subgroup analyses for lesion types. The 95% confidence intervals (CIs) were calculated using the DeLong method with 2000 bootstrapping samples.[29]

## 3. Results

In this VLST, virtual readers assessed paired exams from CT and CXR simulated from a cohort of 294 virtual human subjects. This cohort included 174 individuals with lung nodules and 120 without. The mean size (long axis) of these lesions was 10.09 mm, with a standard deviation of 5.09 mm (Fig. A.2). The smallest lesion measured was 4 mm, while the largest lesion was 34 mm. The lesion sizes were distributed such that 25% of lesions were 6 mm or smaller (first quartile), the median size was 9 mm, and 75% of the lesions (third quartile) were 12 mm or larger. This measurement method follows the approach used in the NLST, though current practice typically averages the two axial dimensions. Validation performances of the virtual reader models for CT and CXR on public datasets are included in

**Appendix C**.

At the patient level, CT outperformed CXR with AUC of 0.92 (95% CI: 0.89-0.95) compared to 0.72 (95% CI: 0.67-0.77). Stratifying the results by lesion type revealed significant ($p < 0.001$) performance differences, but CT consistently outperformed CXR in each category. For the homogeneous lesion type, the AUC for CT was 0.96 (95% CI: 0.94-0.98), substantially higher than for CXR at 0.83 (95% CI: 0.78-0.87) (**Fig. 4**). In contrast, for heterogeneous lesions, both modalities showed lower AUCs: CT 0.89 (95% CI: 0.86-0.92) versus CXR 0.66 (95% CI: 0.60-0.72) (**Fig. 4**). Both NLST and VLST findings reinforce the trend of CT outperforming CXR in lung cancer detection.

## 4. Discussion

The main purpose of this study was to establish a VIT platform capable of emulating certain aspects of major clinical trial in the modalities of CXR and CT imaging. To that end, this Virtual Lung Screening Trial for lung nodule detection represents a preliminary effort to model three fundamental elements of the real-world lung screening trial study: patients, scanners, and readers. By simulating a diverse virtual patient population, utilizing validated radiologic simulators for imaging, and employing machine learning algorithms as standardized virtual readers, our approach provides robust comparison using each virtual subject as their own control, with completely reproducible experimental methods in the end-to-end trial process. Reflecting real-world radiological assessments, CT outperformed CXR in nodule detection.

The composition of the trial population stands as a pivotal element in the execution and success of any clinical study. The VICTRE trial, one of the first published virtual trials in breast imaging, represents a significant milestone within the realm of virtual imaging trials, exhibiting results that are promising when juxtaposed with those derived from human trials.[12] In comparison, our investigation demonstrated the trial effect using a smaller, strategically designed virtual cohort, in contrast to the larger cohorts used in existing clinical lung screening studies.[4,5,7] Despite this, our study extended beyond the scope of prior VIT research, which typically focused on simulating specific pathologies or singular organs.[11,12] The construction of our trial population posed challenges to replicate the complex thoracic anatomy across a spectrum of age, sex, and race/ethnicity to reflect the diverse population encountered in actual clinical trials.[4] Furthermore, we calibrated our virtual models to account for a range of lesion sizes and types. This diversity and clinical realism not only enhanced the relevance of our findings but also facilitated future research in complex bodily system simulations.

An advantage of DukeSim (virtual scanner) over physical scanners is its ability to simulate a wide range of imaging conditions and anatomical variations without the limitations of radiation exposure, patient variability, or scanner availability. This allows for more controlled experiments, rapid prototyping of imaging techniques, and optimization of screening protocols, all in a cost-effective and risk-free virtual environment, which is not feasible with physical scanners. While this study sought to mirror NLST-era scanners, other VITs may evaluate the impact of newer-generation technologies on performance. Using virtual patients also allowed scanning the same patient with both modalities, thus using each patient as their own control to provide greater power, which would not be possible in the real world due to the increase in radiation risk.

The virtual readers were designed using established deep-learning libraries[32] and were trained on publicly available clinical datasets[26,27] to ensure that the development process remained reproducible and broadly applicable. Unlike traditional VITs, which typically employ mathematical observer models that evaluate specific imaging regions for the presence or absence of signal,[12,33] our virtual readers incorporated a search process.[25] This process mirrored the diagnostic approach of a radiologist to first detect and then characterize lesions. In previous VIT studies for COVID-19 detection, we showed that deep-learning models can be susceptible to training and testing biases, which were more pronounced for CXR systems with their great diversity of image appearances.[34-36] To minimize overtraining bias, we have deliberately avoided complex model architectures and elaborate training methodologies, thereby aligning the virtual readers' functionality as closely as possible with that of human readers across different imaging modalities. Reported results demonstrated that CT outperformed CXR in nodule detection, which was consistent with the performance seen in clinical lung cancer screening trials. Our simulated homogenous lesions yielded higher detection performance in both modalities compared to the heterogeneous, demonstrating the effect on trial performance due to lesion texture.

## 5. Limitations:

This study has several limitations. It was very time consuming to generate the complex 3D data, as a result the number of patients was much smaller compared to traditional clinical trials. This study serves as a preliminary demonstration of in silico trial capabilities for lung cancer screening, focusing on patient-level detection of cancer, similar to the recall task of clinical trial radiologists. Despite the small sample, our study design allowed efficient comparisons by assessing detection performance using a heavily enriched sample (>50% cancer prevalence compared to 4% for NLST) and paired imaging of each patient with both modalities. Future work will expand the virtual population and add lesion types such as semi-solid and ground glass. In terms of the virtual reader, only a single deep learning model was developed for each modality, limiting the assessment of inter-observer variability; this can be addressed by incorporating multiple models with varying architectures and training levels. Finally, variations in CXR performance were observed in VLST, potentially due to virtual scanner artifacts or virtual reader performance, highlighting areas for future improvement. For CT, the results demonstrated strong alignment with expected sensitivity, specificity, and AUC, reinforcing its effectiveness over CXR for cancer recall. These findings underscore the potential of VITs to capture key aspects of real-world clinical scenarios. While this study lacks several features of traditional multi-year screening trials, such as long-term endpoints like mortality or confirmed cancer diagnosis, it represents an early and crucial step toward charting a roadmap for comprehensive in-silico screening trials. [37,38] Acknowledging these limitations, future research will focus on enhancing VITs to better replicate clinical trial scenarios, incorporating outcomes such as progression or survival while improving data efficiency.

## 6. Conclusion:

In conclusion, this study presented one of the first end-to-end VITs in the domain of chest imaging, which involve the use of CXR and CT modalities. These modalities are the subject of considerable research interest due to their high clinical volume and impact. Together, the complexity of human models, versatility of scanners, and robustness of readers contribute to the advancement of virtual trials for complex bodily systems and imaging challenges. The transformative potential of virtual imaging trials in advancing evidence-based medicine offers an efficient and ethically conscious approach to medical research and development.

## Data availability

The anthropomorphic phantoms, simulated images, and AI readers may be requested at https://cvit.duke.edu/resources/. Additional project details can be accessed at https://fitushar.github.io/VLST.github.io/.

## Code availability

Code availability DukeSeg and quality control module is publicly available at https://xcat-3.github.io/. Other post-processing modules can be made available upon request for research purposes.

## Acknowledgments

This work was funded in part by the National Institutes of Health (P41-EB028744, R01EB001838, R01HL155293).

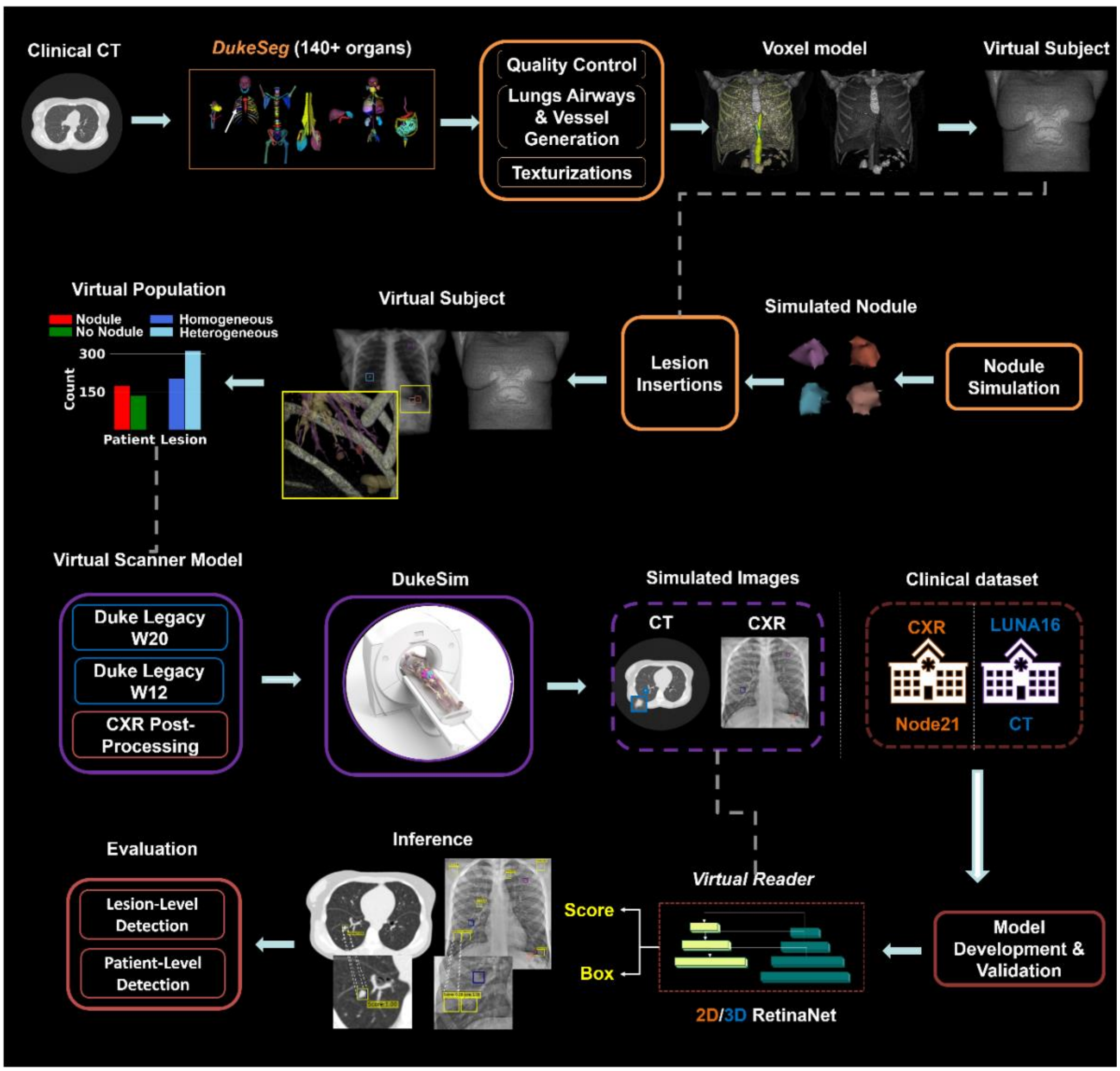

**Figure 1.** Flowchart summarizing the comprehensive workflow of the Virtual Lung Screening Trial (VLST), from creation of virtual patient models using clinical CT data, through lesion simulation and insertion, to virtual imaging using DukeSim, and concluding with evaluation by a virtual reader employing 2D/3D RetinaNet for both lesion-level and patient-level detection evaluation.

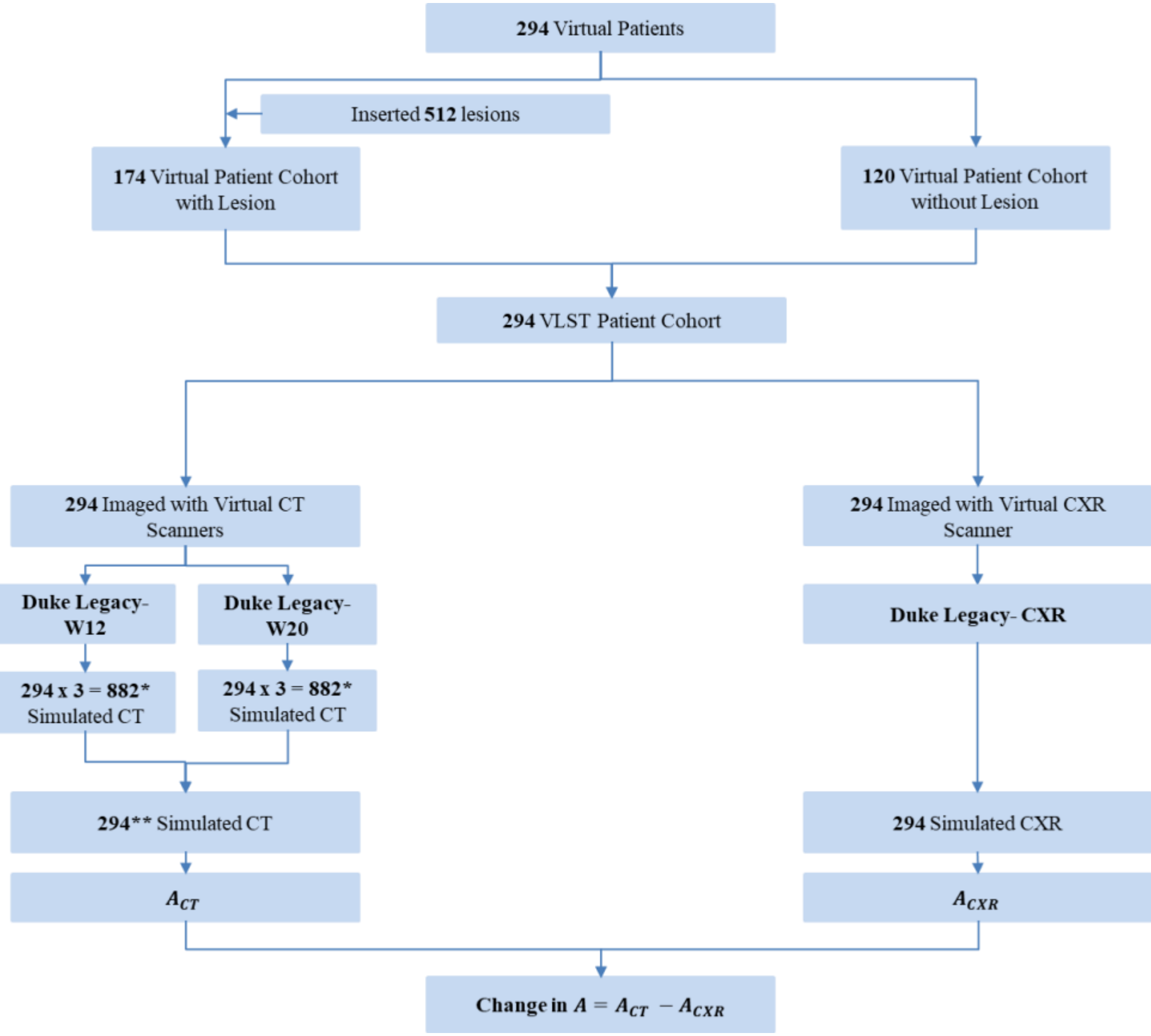

**Figure 2.** Virtual Patient's inclusion and exclusion criteria and progress through the study. Within the study, 294 virtual patients were assessed, with 174 of them having a total of 512 lesions, varying from homogeneous to heterogeneous in nature. All 294 virtual patients underwent both virtual CT and CXR scans. For the CT cohort, *294 virtual images were processed through two distinct scanners, the Duke Legacy-12 and the Duke Legacy-20, with each scanner producing three unique imaging configurations per patient (294 x 3=882). **From these six configurations, one CT image per patient, total 294 was randomly selected for evaluation. As for the CXR cohort, all 294 virtual patients were successfully imaged using the Duke Legacy-CXR scanner. A indicate the area under the receiver operating characteristic curve.

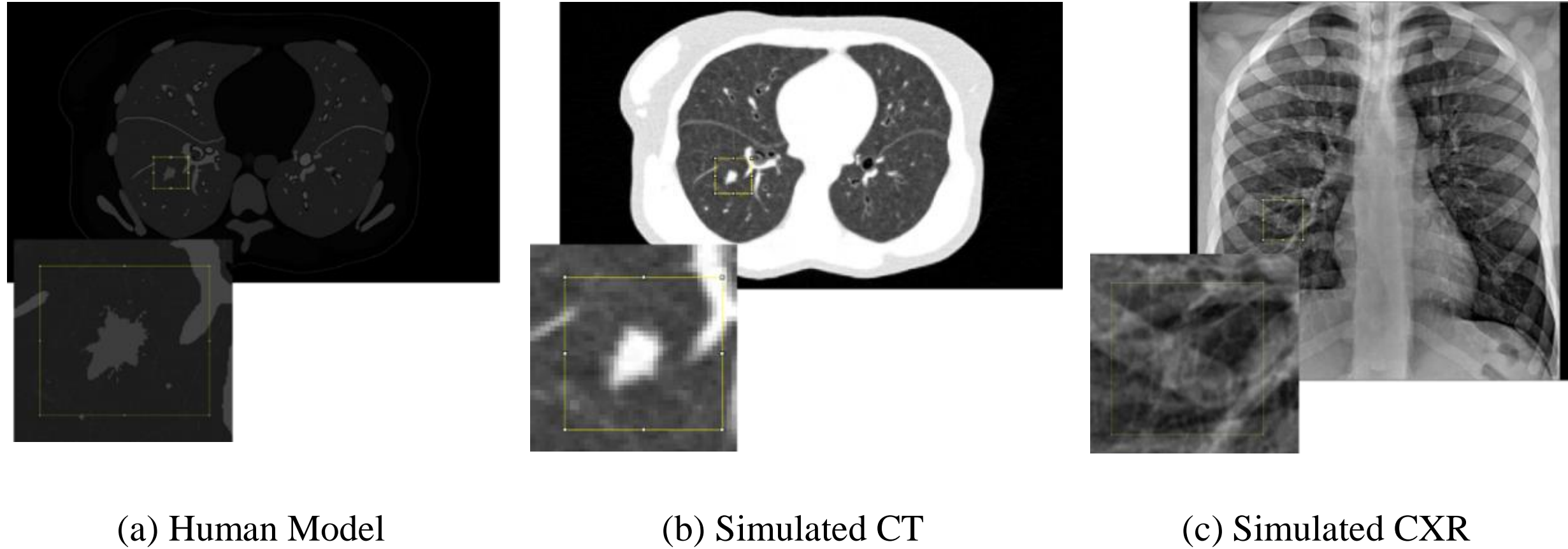

(a) Human Model (b) Simulated CT (c) Simulated CXR

**Figure 3.** Example of human model and simulated images from the Virtual Lungs Screening Trial. Selected slice of (A) computation human model with a homogenous lesion (B) simulated CT scan image from Duke Legacy W20 scanner C) simulated CXR image using legacy post-processing.

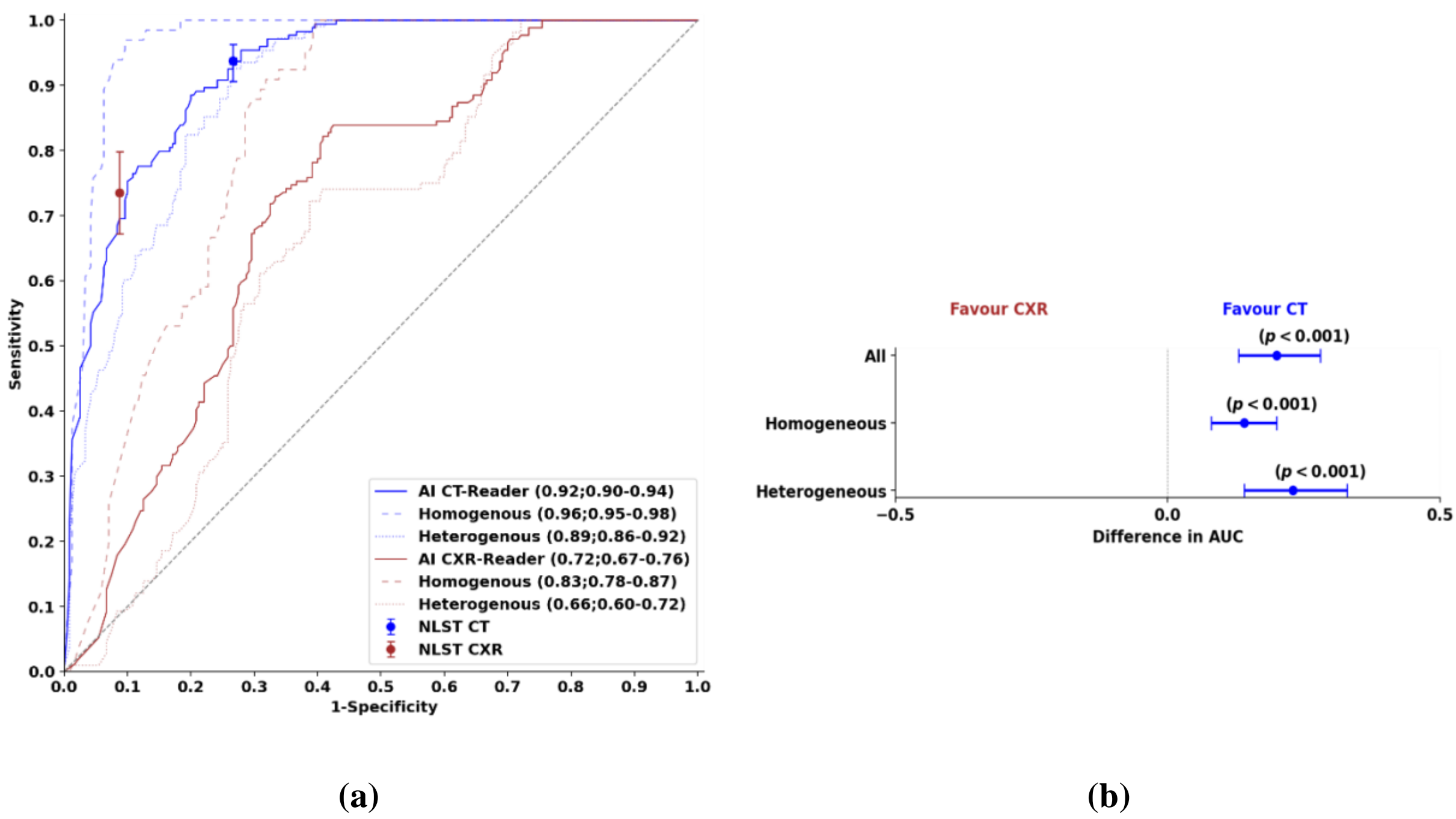

**(a)** **(b)**

**Figure 4.** Performance of the virtual readers predicting lung cancer across patient and lesion types. **(a)** Receiver operating characteristic (ROC) curves comparing the performance of the AI CT-Reader and AI CXR-Reader for patient-level predictions, homogeneous lesions, and heterogeneous lesions. Blue

and brown lines represent AI CT-Reader and AI CXR-Reader performance, respectively. Homogeneous lesions are represented by a solid line, and heterogeneous lesions by a dashed line. The diagonal dashed line represents the ROC curve for a random classifier with an AUROC of 0.50. NLST CT and NLST CXR results are marked with blue and brown dot (error bar represents 95% CI of sensitivity and specificity), respectively. **(b)** Forest plot showing the difference in AUROC between AI CT-Reader and AI CXR-Reader for patient-level predictions. The blue and brown markers represent the AUROC differences for CT and CXR, with error bars indicating the 95% confidence intervals. The p-value for statistical significance is shown ($p < 0.001$).

AI = artificial intelligence. ROC = receiver operating characteristic. AUROC = area under the receiver operating characteristic curve. NLST = National Lung Screening Trial.

**Table 1. Cohort Characteristics of the VLST Virtual Population of With Cases Corresponding to with and without simulated nodule.**

| **Characteristics** | **All** | **With Lung Nodule (n=174)** | **Without Lung Nodule (n=87)** |
|---|---|---|---|
| **Age (year)** | | | |
| Mean ± Std | 59 ± 14 | 59 ± 15 | 60 ± 13 |
| **Sex** | | | |
| Male | 145 (55.56%) | 95 (54.60%) | 50 (57.47%) |
| Female | 116 (44.44%) | 79 (45.40%) | 37 (42.22%) |
| **Weight (kg)** | | | |
| Mean ± Std | 78± 20 | 77 ± 20 | 80 ± 20 |

| BMI | | | |
|---|---|---|---|
| Mean ± Std (min-max) | 27± 6<br>(13-50) | 26± 6<br>(13-50) | 28 ± 5<br>(18-41) |
| **Race** | | | |
| White | 196 (75.10 %) | 128 (73.56%) | 68 (78.16%) |
| Black or African American | 55 (21.07%) | 36 (20.69%) | 19 (21.84%) |
| Other/Unknown | 10 (3.83%) | 10 (5.75 %) | |
| **Ethnic** | | | |
| Not Hispanic or Latino | 257 (98.47%) | 171 (98.28%) | 86 (98.85 %) |
| Hispanic/ Unknown | 4 (1.53 %) | 3 (1.73 %) | 1 (1.15%) |

***Note: Characteristics for the existing 33 without lung nodule human models are currently unavailable.**

# Appendices

**Tushar F.I, et al.** Virtual Lung Screening Trial (VLST): An *In Silico* Study Inspired by the National Lung Screening Trial for Lung Cancer Detection

**Appendix A.** VLST simulated nodules

**Figure 1.** Shown some simulated nodules of different sizes and types imaged with Duke legacy scanners

**Figure 2.** Shown the simulated nodules size (largest axis) distribution

**Appendix B.** VLST imaging physics

**Table 1.** Parameters of Duke Legacy W20 and W12 scanners.

**Appendix C.** VLST Reader

**Appendix C.1.** CT VLST Reader: CT

**Figure 3.** Shown the LUNA16 dataset characteristics

**Figure 4.** Shown the (a) FROC (average sensitive over 1/8, 1/4, 1/2, 1, 2, 4, and 8 FPs per scan) and (b) lesion-level AUC (Area Under the Curve), with 95% confidence interval performance of the CT virtual reader on the LUNA16's (fold-6) training and validation dataset performance

**Appendix C.2.** CXR VLST Reader: CT

**Figure 5.** Shown the lesion-level AUC (Area Under the Curve), with 95% confidence interval performance of the CXR virtual reader on the internal NODE21 test dataset and external VinDr-CXR test dataset performance

**Appendix A: VLST simulated nodules**

Several instances of these simulated nodules, featuring diverse characteristics such as size and morphology, are depicted in **Fig. 1.** The 174 cohort of computational human models are embedded with 512 simulated lung nodules, encompassing both homogeneous (n=202) and heterogeneous (n=310) types with varying sizes, representing a comprehensive range of characteristics for accurate replication of NLST. **Fig. 2** shows the simulated nodules size (largest axis) distribution.

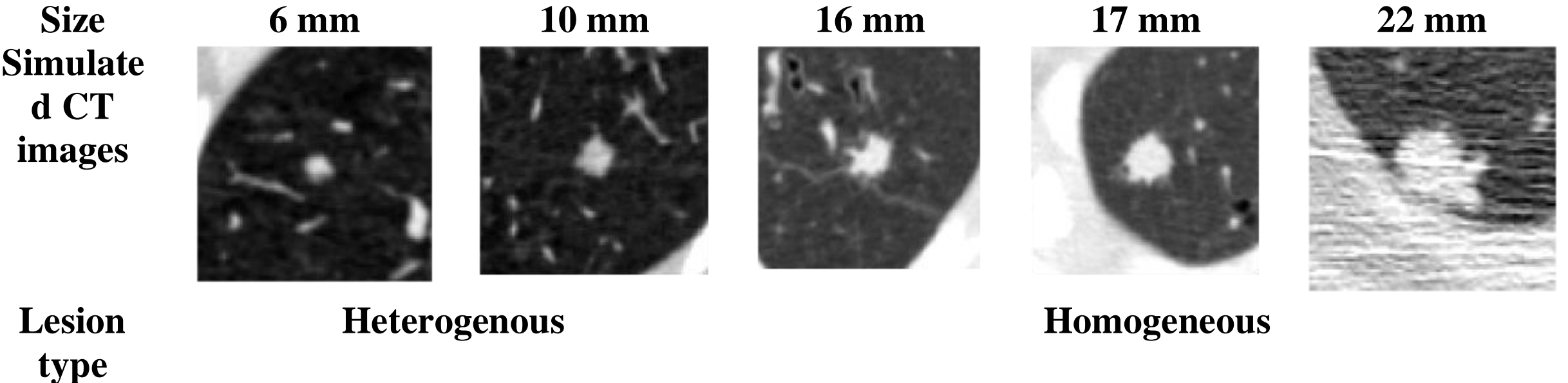

**Figure 1:** Shown some simulated nodules of different sizes and types imaged with Duke legacy scanners.

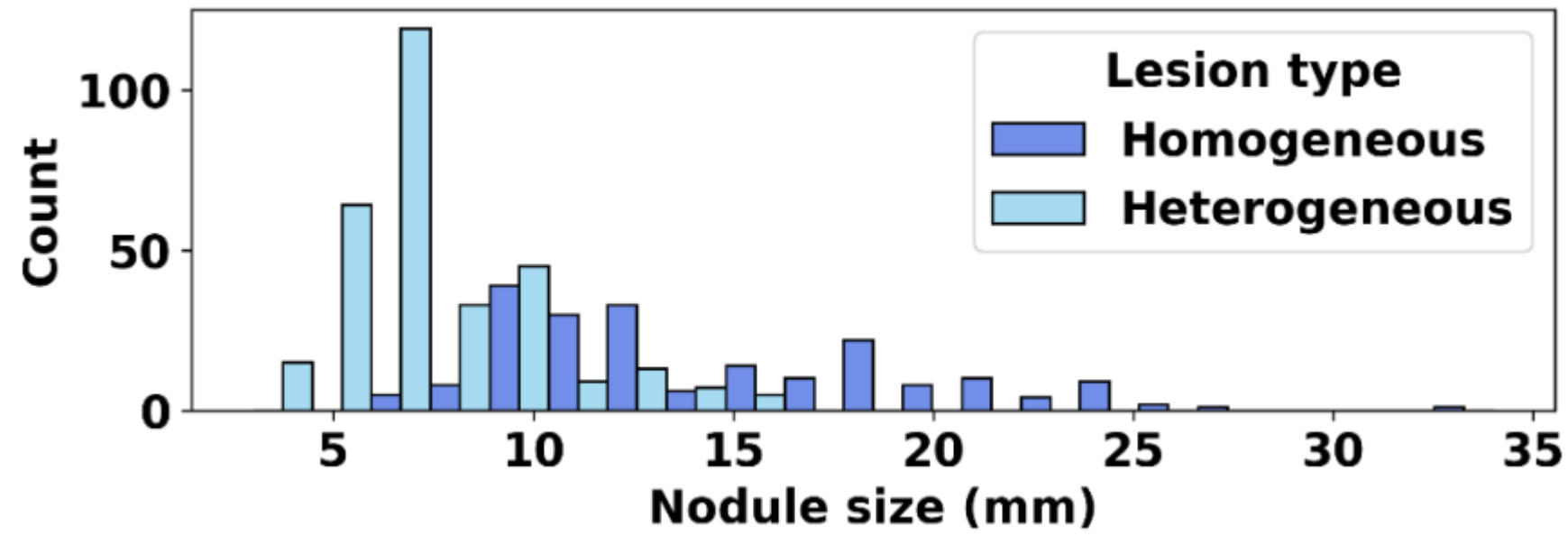

**Figure 2:** Shown the simulated nodules size (largest axis) distribution.

**Appendix B: VLST imaging physics**

**Table 1:** Parameters of Duke Legacy W20 and W12 scanners.

| **Parameters*** | **Duke-Legacy W20** | **Duke-Legacy W12** |
|---|---|---|
| Number of projections per rotation | 1000 | 1000 |
| **Pitch (changeable by user)** | 1.375 | 1.5 |
| **Collimation (mm)** | 20.0 | 12.0 |
| **SIsoD (mm)** | 541 | 570 |
| **SID (mm)** | 949 | 1040 |
| **focal spot (mm)** | 0.7x0.7 | 0.7x0.5 |
| Number of channels | 16 | 16 |
| **Channel width (mm)** | 2.19 | 1.5 |
| **Number of rows** | 900 | 672 |
| **Row width** | 1.0 | 1.5 |
| **Anode angle (degrees)** | 8 | 7 |

* Boldfaced parameters are different between the two scanners

## Appendix C: VLST Reader

Our study incorporates a virtual reader component that simulates the image analysis process usually conducted by radiologists. It utilizes two RetinaNet models—one 2D and one 3D—specifically designed to process CT and CXR images, functioning as virtual readers.

### Appendix C.1: CT VLST Reader

The Virtual Reader for CT scans has undergone development and validation leveraging the LUNA16 clinical dataset. LUNA16,[1] a publicly accessible dataset, encompasses 1,186 radiologist-verified nodule annotations drawn from 600 CT scans.

**Fig. 3** presents a composite visualization, illustrating various aspects of the dataset, including the count of nodules, the distribution of nodule sizes, and the spread of data across different CT scan manufacturers. This dataset is pre-defined in 10 cross-validation folds, where in each fold the model is trained with 9 folds and validated on 1-fold.[1]

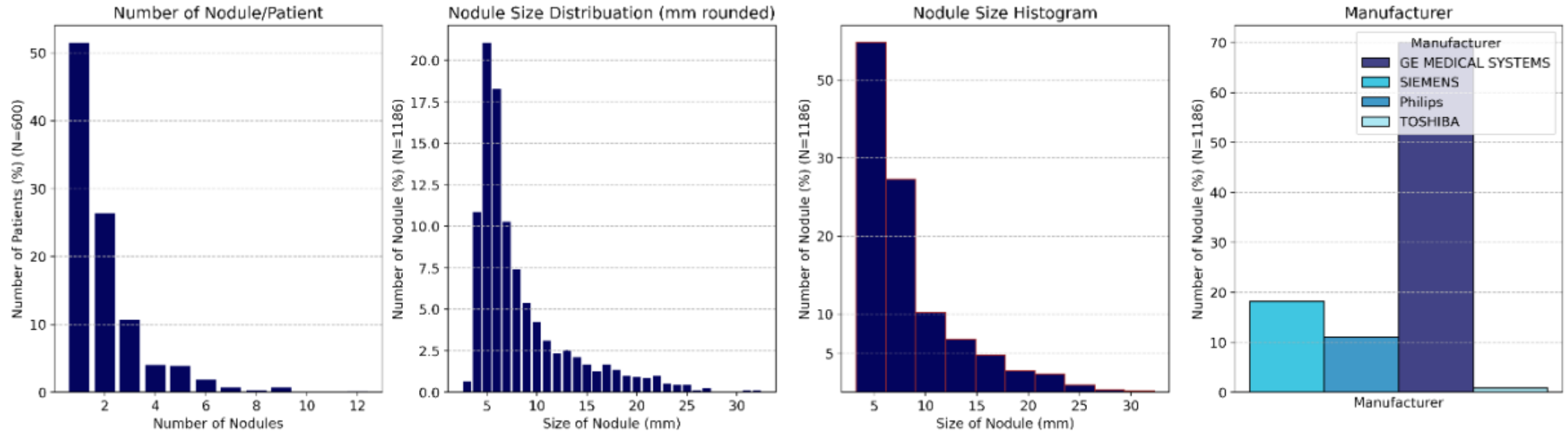

**Figure 3:** Shown the LUNA16 dataset characteristics.

The CT were resampled to $0.7 \times 0.7 \times 1.25$ $(X \times \mathrm{Y} \times \mathrm{Z})$ $mm$ before training and evaluation. The Virtual readers were developed utilization MONAI detection module.[2] CT virtual reader is an 3D detection RetinaNet model. The model was trained using official 10 folds. For each fold, 95% of the training data is used for training, while the rest 5% is used for validation and model selection. The model is trained for 300 epoch.

The model's assessment involved evaluation utilizing the free-response operating characteristic (FROC) analysis method. The ultimate performance metric, denoted as the mean sensitivity at specific predefined false positive rates (FPRs) – specifically, 1/8, 1/4, 1/2, 1, 2, 4, and 8 FPs per scan – was employed. Subsequently, while a significant portion of the models underwent training on 90% of the data, our analytical pursuit concentrated on a single-fold model (fold-6) for consistency and methodological relevance. **Fig. 4** shown the model's training and validation performance.

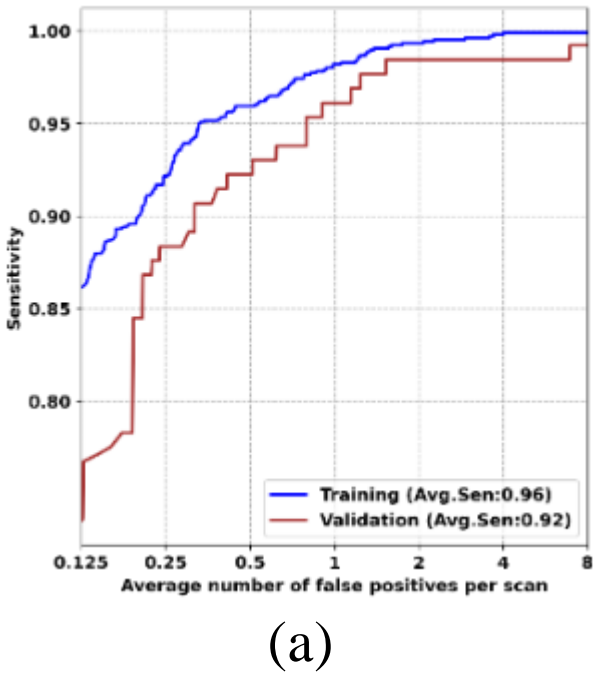

(a)

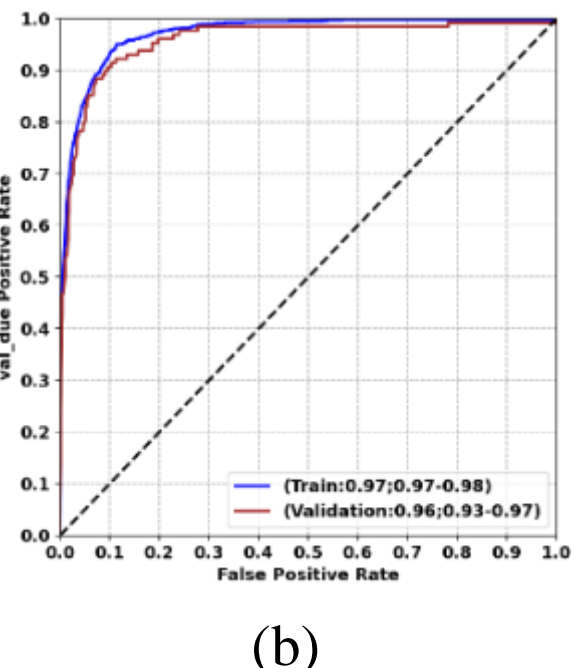

(b)

**Figure 4:** Shown the (a) FROC (average sensitive over 1/8, 1/4, 1/2, 1, 2, 4, and 8 FPs per scan) and (b) lesion-level AUC (Area Under the Curve), with 95% confidence interval performance of the CT virtual reader on the LUNA16's (fold-6) training and validation dataset performance.

**Appendix C.2: CXR VLST Reader**

A 2D RetinaNet-based model has been developed as a virtual reader for lung cancer detection in virtual lung screening trials using chest radiographs (CXR). The NODE21[3] public CXR training dataset consists of 4,882 frontal chest radiographs sourced from well-known repositories, including JSRT,[4] PadChest,[5] ChestX-ray14,[6] and Open-I.[7] Of these, 1,134 images contain 1,476 annotated lung nodules with bounding boxes, representing the positive class, while the remaining 3,748 nodule-free images form the negative class.

For model training, we focused exclusively on the 1,134 images with annotated nodules from the NODE21 dataset. The dataset was split into 80% training and 20% testing, with stratification based on patient origin and data source, ensuring a balanced and representative evaluation. During model development, 80% of the training data was used for training, and 20% of the training set was reserved for validation to select the best-performing model, which was then evaluated on the independent test set. Additionally, model is been evaluated on an external subset of VinDr-CXR dataset.[8] Regarding preprocessing, as recommended by the NODE21 dataset, utilized openCXR library standardization coupled with a specific cropping technique to isolate the lung boundaries in the CXR images.[3] **Fig. 5** shown the lesion-level performance of the CXR virtual readers on the internal and external clinical test datasets.

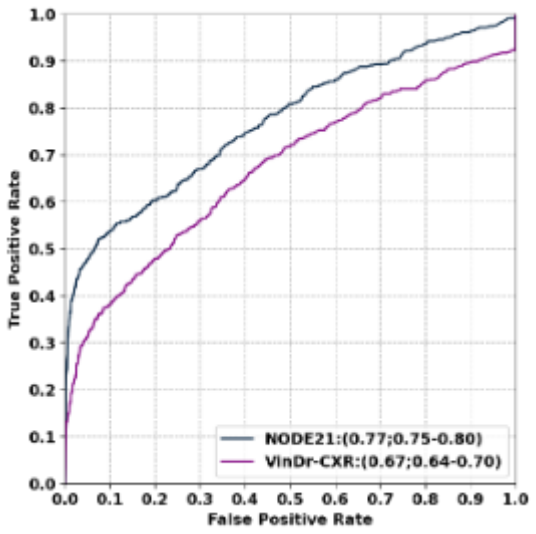

**Figure 5:** Shown the lesion-level AUC (Area Under the Curve), with 95% confidence interval performance of the CXR virtual reader on the internal NODE21 test dataset and external VinDr-CXR test dataset performance.